\documentclass[a4paper,12pt]{article}

\usepackage{psfrag}
\usepackage{graphicx} 
\usepackage{amsmath}
\usepackage{amsthm}
\usepackage{amsfonts}
\usepackage{bm}
\usepackage{amssymb}
\usepackage[]{natbib}

\bibpunct{(}{)}{,}{a}{}{,}

\allowdisplaybreaks[1]

\usepackage{epic} 
\usepackage{ecltree} 
\usepackage{pstricks} 
\usepackage{pst-node} 
\usepackage{psfrag} 

\usepackage{hyperref} 
\hypersetup{
    pdftitle={Full Open Population Capture-Recapture Models with Individual Covariates},
    pdfauthor={\textcopyright Matthew Schofield and Richard Barker}
}

\usepackage{setspace}
\begin{document}

\begin{spacing}{1.9}

\title{Full Open Population Capture-Recapture Models with Individual Covariates}
\author{Matthew R. Schofield\thanks{Department of Statistics, Columbia University, New York, NY, USA} \thanks{Current Address: Department of Statistics, University of Kentucky, Lexington, KY 40506, USA. E-mail: \href{mailto:matthew.schofield@uky.edu}{matthew.schofield@uky.edu}} and Richard J. Barker\thanks{Department of Mathematics and Statistics, University of Otago, PO Box 56, Dunedin, New Zealand. E-mail:  \href{mailto:rbarker@maths.otago.ac.nz}{rbarker@maths.otago.ac.nz}}\\
}
\date{}
\maketitle

\vspace{-1cm}
\begin{abstract}
Traditional analyses of capture-recapture data are based on likelihood functions that explicitly integrate out all missing data.  We use a complete data likelihood (CDL) to show how a wide range of capture-recapture models can be easily fitted using readily available software JAGS/BUGS even when there are individual-specific time-varying covariates.  The models we describe extend those that condition on first capture to include abundance parameters, or parameters related to abundance, such as population size, birth rates or lifetime.  The use of a CDL means that any missing data, including uncertain individual covariates, can be included in models without the need for customized likelihood functions.  This approach also facilitates modeling processes of demographic interest rather than the complexities caused by non-ignorable missing data.  We illustrate using two examples, (i) open population modeling in the presence of a censored time-varying individual covariate in a full robust-design, and (ii) full open population multi-state modeling in the presence of a partially observed categorical variable.
\end{abstract}

\section{Introduction}
An important feature of capture-recapture modeling is the ability to include covariates, including individual-specific ones \cite[]{Lebreton1992,Schwarz1993,Bonner2006,King2008a,Catchpole2008, Bonner2010}.  
The development of methods for including individual covariates has focused on models that condition on the first capture of each individual.  A consequence is that likelihood based inference is restricted to statements about survival or recapture probabilities and related quantities.  Importantly, these models do not include abundance parameters (or parameters related to abundance, such as population growth rates or stopover time) in the likelihood.  Instead, inference about abundance has relied on ad-hoc Horvitz-Thompson-type approaches \cite[]{Huggins1989,McDonald2001}.

Individual-specific time-varying covariates pose problems in models based on the full likelihood as the covariate values cannot be observed for individuals that were available for capture but were not caught.  In order to fit these models by maximum likelihood, we would need to integrate out the missing covariate values for the unseen individuals for each possible value of abundance, as well as integrating out the missing values for the individuals we observed.  
Explicit integration of the likelihood leads to the observed data likelihood (ODL), but this integration is often difficult to do in practice.  An alternative approach is to model in terms of the complete data likelihood (CDL), where specialized computational algorithms, such as the Gibbs sampler or the EM algorithm perform the required integration during the model fitting process.  \cite{Schofield2007,Schofield2008,Schofield2009} lay out a unified framework for capture-recapture modeling using the CDL, which while flexible enough to include individual-specific time-varying covariates, still requires the user to construct these algorithms in order to fit the model.  The same process is required when including individual covariates in closed population studies, for example, see \cite{King2008, Royle2009}.

To date, no user-friendly software is available for full open population modeling in the presence of individual-specific time-varying continuous covariates.  Algorithms do exist for other individual-specific open population models, including user-written algorithms for the multi-state model \cite[]{Dupuis2007} and models where survival and fecundity parameters depend on population abundance \cite[]{Schofield2008}.  Unfortunately, these algorithms require custom-written software and lack generalizability.  Another approach is to include these models in JAGS \cite[]{Plummer2003} or BUGS \cite[]{Lunn2000}, software that fits models using the Gibbs sampler and is being increasingly used by ecologists \cite{Royle2007,Gimenez2009,Schofield2009a,Link2010}.  This has been done for individual-specific mixed effects models in \cite{Royle2008,Link2010}.

Here we consider models for two datasets, both of which have individual-specific time-varying covariates.  The first is an example from \cite{Nichols1992}, where the robust design is used to sample meadow voles, \textit{Microtus pennsylvanicus}, with the body mass measured every time there was a capture.  The body mass measurements were discretized in order to fit a multi-state model, conditional on first capture, to understand how the covariate body mass changed through time.  This was later refined by \cite{Bonner2006} who model body mass as an individual-specific time-varying covariate, although, their analysis ignored the robust design and conditioned on first capture.  Other analyses have used the robust design to estimate abundance, but did this without taking into account body mass \cite[Pg. 525]{Williams2002}.  Here we use information on the body mass within the robust design to estimate abundance.

The second dataset is a study of conjunctivitis in house finch, \textit{Carpodacus mexicanus}, where the covariate of interest is disease status, a covariate that is not always observed.  \cite{Conn2009} model the data using an extension to the multi-state model to account for the partially observed data, however, they only model conditional on first capture.  Even though they are able to obtain estimates of survival for each disease class and movement between the two disease classes, they were unable to get an predictions of the population in each disease class through time.

Here we show how the framework of \cite{Schofield2007,Schofield2008,Schofield2009} can be used to extend the models of \cite{Royle2007}, \cite{Royle2008} and \cite{Link2010} to fit full open population models with individual-specific time-varying covariates in BUGS.   There are only minor differences in the model and BUGS code for the two examples, despite the differences between the  datasets: (i) continuous vs. categorical covariate, (ii) robust design vs. standard capture-recapture design, (iii) covariate measurement every capture occasion vs.  uncertainty in covariate measurements on some capture occasions.  This shows the power and flexibility of the modeling approach, since the two examples include as special cases: the multi-state model, multi-event type models, individual-specific time-varying continuous covariates, and lifetime duration models such as the stop-over model.

\section{Model Framework}\label{sect:frame}
Capture-recapture models involve complex missing data mechanisms.  Traditional approaches to inference focus on deriving the likelihood for the observed data (the ODL) by integrating over all missing data.  Instead, we use the modeling framework of \cite{Schofield2007,Schofield2008,Schofield2009} that uses data augmentation \cite[]{Tanner1987} to allow us to model in terms of the complete data likelihood (CDL).  Similar ideas have also been proposed by \cite{Royle2008}.

The likelihood we use for inference is in terms of the complete data, which for a capture-recapture study with individual-specific time-varying covariate data are the (i) times of birth, (ii) times of death, and (iii) complete covariate values for each individual ever available for capture.  The main advantage of using this likelihood over the ODL is that we are able to focus on modeling the processes of interest rather than having to account for the complexities caused by missing data that result from sampling methods.  
Importantly, in adopting the CDL approach to inference, we do not need to make any additional assumptions to those made when using the ODL.  It is simply a reformulation of the model in terms of the easier-to-understand CDL where we use computational algorithms, such as Markov chain Monte Carlo (MCMC) or the expectation-maximization (EM) algorithm, to integrate over all missing data. 

We write the CDL for the capture-recapture model including birth and covariates as
\begin{equation}\label{eq:cmascov}
  \underbrace{[a^{b}\vert \theta^{b},N]}_{\text{Birth}}\underbrace{[a^{d}\vert a^{b},\theta^{d},N]}_{\text{Mortality}}\underbrace{[X\vert a^{b},a^{d},\theta^{X},N]}_{\text{Capture}}\underbrace{[z\vert \theta^{z},N]}_{\text{Covariate}},
\end{equation}
\begin{description}
 \item where $[Y|X]$ is the probability (density) of the random variable $Y$ given $X$;
 \item $a^{b}_{ij}=0$ means that individual $i$ has yet to be born at, or before, sampling period $j$, with $a^{b}_{ij}=1$ otherwise;
 \item $a^{d}_{ij}=1$ means that individual $i$ has yet to die at sampling period $j$, with $a^{d}_{ij}=0$ otherwise;
 \item $X_{ij}=1$ means that individual $i$ was caught in sampling period $j$, with $X_{ij}=0$ otherwise;
 \item $z_{ij}$ is the covariate value for individual $i$ in sampling period $j$;
\end{description}
$\theta^{b}$ are parameters describing the birth process, $\theta^{d}$ are parameters describing the mortality process, $\theta^{X}$ are parameters describing the capture process, $\theta^{z}$ are parameters of the covariate distribution and $N$ is the total number of individuals available for capture during the study period (which is distinct from $N_{j}$, the number of individuals alive in the $j$th sampling occasion).
The covariate $z$ can be used to help model the birth, death and covariate processes, although we assume that this is specified through the models for $\theta^{b}$, $\theta^{d}$ and $\theta^{X}$.  In order for inference to be valid, we must ensure that these models do not violate the laws of conditional probability, for example, by assuming that survival probability depends on the covariate value at the end of the period.

Most demographic summaries of interest are obtained as derived quantities of $a^{d}$ and $a^{b}$.  For example, the population size in each sampling period, $N_{j}$, and the ``lifetime'' for each individual $\Delta_{i}$ are
\begin{align*}
   N_{j} &= \sum_{i=1}^{N}a^{b}_{ij}a^{d}_{ij},~~j=1,\ldots,k,\\
   \Delta_{i} &= \sum_{j=1}^{k}a^{b}_{ij}a^{d}_{ij},~~i=1,\ldots,N.
\end{align*}
Other potential quantities of interest we could specify include the number of births and deaths between each sampling period.

The choice of model for each of the components in (\ref{eq:cmascov}) will depend on the data, and the assumptions we are willing to make.  For a standard capture-recapture study design, we present some common models for each component.  We leave the parameters for mortality and capture to be individual specific as these are being modeled in terms of individual-specific covariates.

\subsection{The Birth Component}
One possible model for the birth components is
\begin{align}\nonumber
    [a^{b}\vert \theta^{b},N] &= \prod_{i=1}^{N}[a^{b}_{i1}\vert\theta^{b}]\prod_{j=2}^{k}[a^{b}_{ij}\vert a^{b}_{i1},\ldots,a^{b}_{ij-1},\theta^{b}]\\\label{eq:condbirth}
    &=\prod_{i=1}^{N}Bern(\zeta_{1})\prod_{j=2}^{k}Bern\left(\prod_{h=1}^{j-1}(1-a^{b}_{ih})\zeta_{j} + \left(1-\prod_{h=1}^{j-1}(1-a^{b}_{ih})\right)\right),
\end{align}
where $Bern(p)$ denotes a Bernoulli distribution with parameter $p$ and we set $\zeta_{k}$ = 1.  We include the term $\prod_{h=1}^{j-1}(1-a^{b}_{ih})\zeta_{j} + 1-\prod_{h=1}^{j-1}(1-a^{b}_{ih})$ to ensure that an individual can only be born once.
The value $\zeta_{1}$ is the probability of being born before the study began, with the values $\zeta_{j+1}$ defined as the probability of being born between sample $j$ and $j+1$ conditional on (i) not being born before $j$, and (ii) being at risk of capture at some point during the study.  A possible reparameterization is
\[  
    \beta_{0} = \zeta_{1},~~ \beta_{j} = \zeta_{j+1}\left(1-\sum_{h=1}^{j-1}\beta_{h}\right),~~j=1,\ldots,k-1.
\]
This gives the birth formulation of \cite{Schwarz1996} where $\beta_{j}$ is the probability of being born between sample $j$ and $j+1$ conditional on being at risk of capture at some point during the study.  Neither $\zeta_{j}$ nor $\beta_{j}$ are meaningful birth parameters, since they are defined in terms of the study/sampling process, as for example, a change in $k$ changes the parameter values.  A more natural parameterization is to use per-capita birth rates,
\[
  \eta_{j} = \frac{\beta_{j}N}{N_{j}},~~j=1,\ldots,k-1.
\]
This gives the birth formulation used in \cite{Schofield2008}, where $\eta_{j}$ is the expected number of births between sampling period $j$ and $j+1$ for each individual in the  population at sampling period $j$ conditional on $N$. These parameters are similar to the birth parameters, $f_{j}$, used by \cite{Pradel1996,Link2005}; the difference is that the denominator they use is $\text{E}[N_{j}|N]$ instead of $N_{j}$.  

\subsection{The Mortality Component}
We factor the component for mortality as,
\begin{align}\nonumber
    [a^{d}\vert a^{b},\theta^{d},N] &= \prod_{i=1}^{N}\prod_{j=2}^{k}[a^{d}_{ij}\vert a^{d}_{ij-1},a^{b}_{ij-1},\theta^{d}]\\\label{eq:conddeath}        
    &=\prod_{i=1}^{N}\prod_{j=2}^{k}Bern(a^{d}_{ij-1}(a^{b}_{ij-1}S_{ij-1} + (1-a^{b}_{ij-1}))),   
\end{align}
where the parameter $S_{ij}$ is the probability of individual $i$ surviving between sampling period $j$ and $j+1$.  The term $a^{d}_{ij-1}(a^{b}_{ij-1}S_{ij} + (1-a^{b}_{ij-1}))$ is required to ensure that an individual can only (i) die after being born, and (ii) live once.  

\subsection{The Capture Component}
We factor the component for capture as,
\begin{align}\nonumber
    [X\vert a^{b},a^{d},\theta^{X},N] &= \prod_{i=1}^{N}\prod_{j=1}^{k}[X_{ij}\vert a^{b}_{ij},a^{d}_{ij},\theta^{X}]\\\label{eq:condcapt}
    &= \prod_{i=1}^{N}\prod_{j=1}^{k}Bern(a^{d}_{ij}a^{b}_{ij}p_{ij}),
\end{align}
where $p_{ij}$ is the probability of capture for individual $i$ in sampling period $j$.  The term $a^{d}_{ij}a^{b}_{ij}$ is required to ensure that an individual is only available for capture while it is alive. 

\subsection{Additions Required For BUGS/JAGS}
In order to make inference about the parameters in the model, we adopt a Bayesian approach and fit all examples using the software JAGS \cite[]{Plummer2003}, with model, data, initial values and script files available at \url{www.maths.otago.ac.nz/~rbarker/BUGS}.   We use JAGS for the following examples due to superior convergence in trial runs of the algorithms as compared to OpenBUGS.  
The modeling language of JAGS is nearly identical to BUGS \cite[]{Lunn2000} and is able to be called from R \cite[]{Su2009}.  We describe the main differences between BUGS and JAGS and how to specify the data and the initial values in the supplementary materials.

Neither JAGS nor BUGS allow stochastic indices, such as $N$, as was used in \cite{Schofield2008}.  
Instead, we must use a computational trick to include $N$.  The trick is given in \cite{Durban2005} and involves specifying $M$, an upper bound for $N$.  
An alternate, yet mathematically equivalent approach is given in \cite{Royle2007}.  They also specify $M$ but then reparameterize the model in terms of an incomplete indicator variable, $w$, instead of $N$.  The value $w_{i}=1$ means that individual $i$ was at risk of capture during the study and $w_{i}=0$ otherwise, with $\sum_{i=1}^{M}w_{i}=N$.
Having used both approaches, we find them practically equivalent with the exception that the specification of \cite{Royle2007} no longer has $N$ available for hierarchical modeling, but does generally run slightly faster than the approach of \cite{Durban2005}.  Since we have no desire to include a hierarchical model for $N$ in the examples we explore, we use the approach of \cite{Royle2007} to make use of the faster algorithm.   For readability, we do not change the CDL in (\ref{eq:cmascov}) to reflect this additional likelihood component and continue to write the CDL conditioning on $N$.


Another difficulty is that attempting to use the per-capita birth rate ($\eta_{j}$) parameterization in JAGS or BUGS results in code that is currently impractically slow to run.  Here we use the less natural $\zeta_{j}$ parameterization described above.  This is not a limitation of MCMC or the Gibbs sampler, as \cite{Schofield2008} used the per-capita birth rates.  Instead it is a current limitation of JAGS and BUGS.  

\section{Example: Meadow Voles}\label{sect:vole}
\cite{Nichols1992} report a capture-recapture study of meadow voles, \textit{Microtus pennsylvanicus}, using a robust design \cite[]{Pollock1982} with $173$ individuals caught over $6$ primary periods each with $5$ secondary samples.  Every time an individual was caught the mass of the animal was recorded to the nearest integer.  The scales used to measure mass had a maximum at $60$ grams, with many individuals censored. 

\cite{Bonner2006} collapsed the data across the secondary periods, and allocated a single measurement to the observed body mass.  Using these data they extended the Cormack-Jolly-Seber model (CJS) to include an individual-specific time-varying continuous covariate.  \cite{Schofield2009a} used the same data to show how to fit this model using BUGS (neither \cite{Bonner2006} nor \cite{Schofield2009a} accounted for the censored data).  Here we extend this model to include the full robust design and make use of all information on body mass.  Including the birth process in the model allows us to estimate the population size of the meadow vole in the presence of the individual-specific time-varying continuous covariate.


The CDL is given in (\ref{eq:cmascov}), with the birth component given in (\ref{eq:condbirth}) and the mortality component given in (\ref{eq:conddeath}).  
To include the robust design, we redefine $X$ to be an array with $X_{ijl}=1$ if individual $i$ is caught in primary period $j$ and secondary period $l$.  We extend the capture component in (\ref{eq:condcapt}) to include both the $k_{1}$ primary periods and the $k_{2j}$ secondary periods,
\begin{align*}
  [X\vert a^{b},a^{d},\theta^{X},N] &= \prod_{i=1}^{N}\prod_{j=1}^{k_{1}}\prod_{l=1}^{k_{2j}}[X_{ijl}\vert a^{b}_{ij},a^{d}_{ij},\theta^{X}]\\
    &= \prod_{i=1}^{N}\prod_{j=1}^{k_{1}}\prod_{l=1}^{k_{2j}}Bern(a^{d}_{ij}a^{b}_{ij}p_{ijl}),
\end{align*}

We also redefine $z$ to be an array with $z_{ijl}$ being the body mass for individual $i$ in primary period $j$ and secondary period $l$.  Since every observed body mass value was either censored or rounded we treat $z_{ijl}$ as missing for every individual in every sampling period and denote the observed masses by $z_{ijl}^{obs}$.  We specify the model for $z$ as
\begin{align*}
  [z\vert \theta^{z},N] &= \prod_{i=1}^{N}\prod_{j=b_{i}}^{k_{1}}\prod_{l=1}^{k_{2j}}[z_{ijl}\vert \theta^{z}]\\
  & = \prod_{i=1}^{N}\prod_{j=f_{i}}^{k_{1}}\prod_{l=1}^{k_{2j}}N(\lambda_{ij},\sigma^{2}_{z})I(\delta_{1ijl},\delta_{2ijl}),  
\end{align*}
where $N(\mu,\sigma^{2})$ denotes a Normal distribution with mean $\mu$ and variance $\sigma^{2}$, $b_{i}$ is the first sample individual $i$ was alive and $I()$ is used to include the rounding, censoring and truncation, with  
\[
   \delta_{1ijl} = \left\{\begin{array}{ll}
                           z^{obs}_{ijl}-0.5 & \text{if $X_{ijl}=1$}\\
                           0 & \text{otherwise}\\
                          \end{array}\right.,~~
   \delta_{2ijl} = \left\{\begin{array}{ll}
                           z^{obs}_{ijl}+0.5 & \text{if $X_{ijl}=1$ and $z^{obs}_{ijl}\neq 60$}\\
                           \infty & \text{otherwise.}\\                           
                          \end{array}\right.
\]
In other words, we assume that during the secondary periods when the population is assumed to be closed, mass remains constant and any differences are attributable to the measurement error $\sigma^{2}_{z}$.  We follow \cite{Bonner2006} and model $\lambda_{ij}$ as
\begin{align*}
\lambda_{ib_{i}} &\sim N(\mu_{\lambda},\sigma^{2}_{\lambda1}), ~~    \lambda_{ij} \sim N(\lambda_{ij-1}+\Delta_{j-1},\sigma^{2}_{\lambda2}),~~j=b_{i}+1,\ldots,k_{1}.
\end{align*}
This is a random walk with drift, where the mass of each individual increases, on average, $\Delta_{j}$ grams between primary period $j$ and $j+1$.


We model parameters $S$ and $p$ as
\begin{gather*}
   \text{logit}(S_{ij}) = \alpha_{0} + \alpha_{1}\lambda^{\prime}_{ij} + \eta^{S}_{j},~~ \eta^{S}_{j} \sim N(0,\sigma^{2}_{S}),\\
   \text{logit}(p_{ijl}) = \gamma_{0} + \gamma_{1}\lambda^{\prime}_{ij} + \eta^{p}_{j} + \epsilon^{p}_{jl},~~ \eta^{p}_{j} \sim N(0,\sigma^{2}_{p1}), ~~ \epsilon^{p}_{jl} \sim N(0,\sigma^{2}_{p2}),
\end{gather*}
where $\lambda^{\prime}_{ij}$ is an approximately standardized value of $\lambda_{ij}$.
We allow individuals to have survival probabilities that depend on their mass, with  additional temporal variability, modeled as a random effect.  
We allow probability of capture to depend on body mass due to allow for a sampling strategy that discriminated due to body mass, with additional variability within the secondary period and between primary periods, both modeled as a random effect.  This is equivalent to specifying the closed population model $M_{t}$ as a random effect, with the mean varying between the primary periods.
 The priors for all parameters are given in the supplementary materials.

In order to determine the effect that rounding and censoring have on the results we also fit the model with $z_{ijl}=z^{obs}_{ijl}$ when $X_{ijl}=1$.

Each of the models was run in JAGS with an adaptive phase of $5000$ iterations followed by a posterior sample of $20000$ iterations.  To ensure convergence we run $3$ parallel chains with different starting values and checked convergence with the Brooks-Gelman-Rubin diagnostic \cite[]{Brooks1998a}.  We combined the posterior samples from the three chains to give a total posterior sample of $60000$.

\begin{figure}[htpb]
  \begin{minipage}{\textwidth}
    \begin{center}
      \psfrag{aaa1}[t][t]{$\mu_{\lambda}$}
      \psfrag{aaa2}[t][t]{$\Delta_{1}$}
      \psfrag{aaa3}[t][t]{$\Delta_{2}$}
      \psfrag{aaa4}[t][t]{$\Delta_{3}$}
      \psfrag{aaa5}[t][t]{$\Delta_{4}$}
      \psfrag{aaa6}[t][t]{$\Delta_{5}$}
      \psfrag{ccc1}[t][t]{$\sigma_{S}$}
      \psfrag{ccc2}[t][t]{$\sigma_{p_1}$}
      \psfrag{ccc3}[t][t]{$\sigma_{p_2}$}
      \psfrag{ddd1}[t][t]{$\sigma_{Z}$}
      \psfrag{ddd2}[t][t]{$\sigma_{\lambda_1}$}
      \psfrag{ddd3}[t][t]{$\sigma_{\lambda_2}$}
      \includegraphics[angle=270,width = \textwidth]{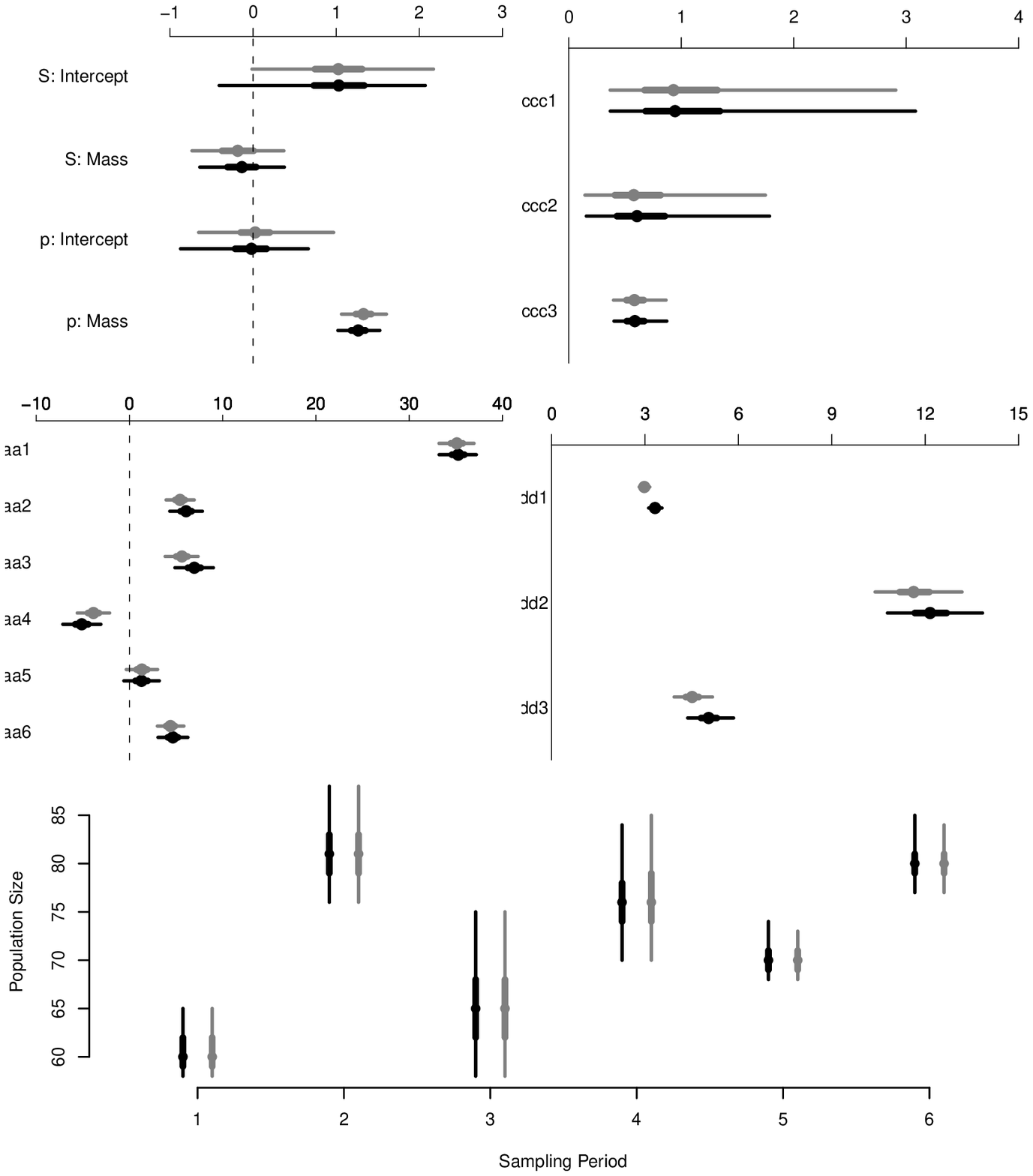} 
      \caption{Parameter estimates for the meadow vole example.  The point gives the median of the marginal posterior distribution and the lines represent the central $50\%$ and $95\%$ credible intervals.  In all plots, black is the model with censoring and blue is the model without censoring.}
      \label{fig:vole}       
    \end{center}
  \end{minipage}
\end{figure}

The results suggest that there is little practical difference between accounting for the censoring of the covariate values and simply modeling using the raw observations (figure \ref{fig:vole}).  
While there are differences in the model for mass, particularly in the variances, this does not translate to substantial differences in the model for survival or probability of capture. 
While mass appeared to be associated with the probability of capture, with larger animals having a higher chance of capture, there is no evidence of body mass being associated with survival.  This result differs from \cite{Bonner2006} and \cite{Schofield2009a} who found that body mass was not associated with either capture probability or survival.  This difference appears to be due to \cite{Bonner2006} compressing the robust design into a more standard CJS design.  To ensure that this is consistent with the data we compared the observed data for individuals caught at least once in any given secondary period, with a significant increase in the average body mass as the number of captures increases.  After adjusting for the effect of body mass on capture probability, the abundance of the meadow vole appears stable, fluctuating between $\sim55$ to $\sim85$ individuals during the study (figure \ref{fig:vole}).

\section{Example: Conjunctivitis in House Finch}\label{sect:finch}
\cite{Conn2009} used a multi-state model to study conjunctivitis in house finch, \textit{Carpodacus mexicanus}, with $813$ individuals caught in $16$ samples.  
A two-state model (whether or not an individual had conjunctivitis) was used, with some individuals having unknown status.  Our approach is to include all missing disease information  using data augmentation and treating the disease as a individual-specific time-varying categorical covariate \cite[]{Dupuis1995}.  Thus we can use the CDL in (\ref{eq:cmascov}) with disease being the covariate $z$.  
Since the covariate takes two values, we can examine abundance, or any other demographic summary, separately for each group.

\cite{Dupuis2007} used the CDL to fit a multi-state model and estimate abundance.  Their approach differs to ours due to different computational algorithms.  To improve the efficiency of their MCMC sampling, they summed over the latent state variables $z_{ij}$ when updating $N$.  While this will yield a quicker algorithm, both in terms of mixing and time, it lacks generalizability beyond the multi-state model to, say, individual-specific time-vary continuous covariates.  In contrast, our approach allows us to apply the CDL across a range of different models and different covariate distributions, including continuous covariates, without major modifications in the JAGS/BUGS code.  

Since many individuals have uncertain disease status, even when encountered, we must consider assumptions about the missingness of these observations \cite[]{Rubin1976}.  Here we assume that they are either missing completely at random or missing at random.  In either case, the process that describes how the data go missing does not need to be included in the model.  In the supplementary materials we describe and fit the model where we assume that the additional missing data is missing not at random.  The results are very similar to those from the model assuming the additional missing data is missing at random.

The CDL is given in (\ref{eq:cmascov}) and we specify the birth component as in (\ref{eq:condbirth}), the death component as in (\ref{eq:conddeath}) and the capture component as in (\ref{eq:condcapt}).  The only component left to specify is the covariate, disease.

We specify the model for disease as
\begin{align*}
  [z\vert \theta^{z},N] &= \prod_{i=1}^{N}[z_{ib_{i}}\vert\theta^{z}]\prod_{j=b_{i}+1}^{k}[z_{ij}\vert z_{ij-1},\theta^{z}]\\
    &=  \prod_{i=1}^{N}Cat((\nu_{1},\nu_{2}))\prod_{j=b_{i}+1}^{k}Cat((\omega_{z_{ij-1}1},\omega_{z_{ij-1}2}))
\end{align*}
where $Cat(\bm{\pi})$ is a categorical distribution with probability vector $\bm{\pi}$.  The covariate value $z_{ij}=2$ indicates that individual $i$ does has the disease in sample $j$ and $z_{ij}=1$ otherwise, $\nu_{l}$ is the probability of being in state $l$ in the first sample after birth and $\omega_{hl}$ is the probability of moving from state $h$ to state $l$.  We use a generic categorical distribution (instead of the binomial) to show how this model generalizes to more than two states.

To complete the model specification we model $S$ and $p$ as
\begin{align*}
   \text{logit}(S_{ij}) &= \alpha_{0} + \alpha_{1}I(z_{ij}=2) + \eta^{S}_{j},~~\eta^{S}_{j} \sim N(0,\sigma^{2}_{S}),\\
   \text{logit}(p_{ij}) &= \gamma_{0} + \gamma_{1}I(z_{ij}=2) + \eta^{p}_{j},~~\eta^{p}_{j} \sim N(0,\sigma^{2}_{p}).      
\end{align*}
We allow individuals to have survival and capture probabilities that depend on their disease status, with temporal variability modeled as a random effect.    The priors for all parameters are given in the supplementary materials.

Each of the models was run in JAGS with an adaptive phase of $25000$ iterations followed by a posterior sample of $25000$ iterations.  To ensure convergence we run $3$ parallel chains with different starting values and checked convergence with the Brooks-Gelman-Rubin diagnostic \cite[]{Brooks1998a}.  We combined the posterior samples from the three chains to give a total posterior sample of $75000$.

\begin{figure}[htpb]
  \begin{minipage}{\textwidth}
    \begin{center}
      \psfrag{aaa1}[t][t]{$\sigma_{S}$}
      \psfrag{aaa2}[t][t]{$\sigma_{p}$}
      \psfrag{ccc1}[t][t]{$\omega_{11}$}
      \psfrag{ccc2}[t][t]{$\omega_{22}$}
      \includegraphics[angle=270,width = \textwidth]{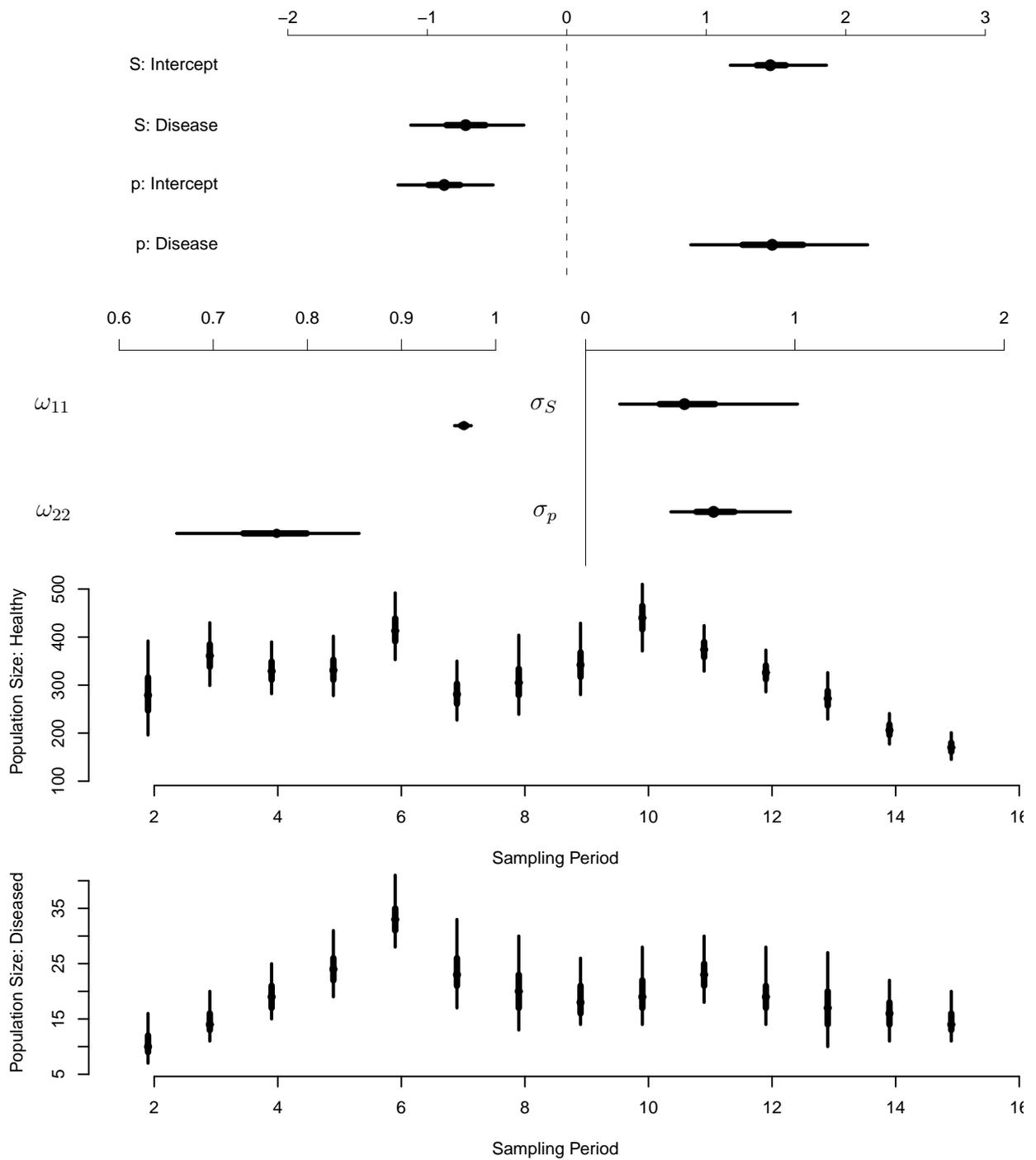} 
      \caption{Parameter estimates for the house finch example.  The point gives the median of the marginal posterior distribution and the lines represent the central $50\%$ and $95\%$ credible intervals.}
      \label{fig:finch}       
    \end{center}
  \end{minipage}
\end{figure}

The results suggest that disease status is associated with both survival and capture probability (figure \ref{fig:finch}).  Having conjunctivitis lowered the log-odds of survival, while increasing the log-odds of capture.  The results for survival appear to agree with \cite{Conn2009}, with no results for capture probability available to compare.  
The transition probabilities suggest that it is rare for an individual to develop conjunctivitis, with less than $5\%$ of disease free animals contracting the disease (figure \ref{fig:finch}).  However, once an individual has conjunctivitis it is somewhat difficult to become disease free, with around three quarters of individuals remaining in the disease state from one sample to the next.
The population size for diseased animals, while low, appears to be relatively stable with no fewer than $5$ diseased individuals in the population (and as many as $40$) during the study (figure \ref{fig:finch}).  It is interesting to note that the population sizes for the diseased and non-diseased states, while similar, do not necessarily exhibit the same dynamics through time.  In particular, one could claim that there are periods where one class increases or decreases while the other remains relatively stable.

\section{Discussion}
We have shown how to use the framework of \cite{Schofield2007,Schofield2008,Schofield2009} to fit using JAGS or BUGS complex models where we estimate demographic summaries of interest in the presence of individual-specific time-varying covariates.  The two examples we show include modeling in the presence of covariate uncertainty and including different study designs.
The CDL conveniently factorizes so that we are able to specify separate models for the birth, mortality, capture and covariate processes while fitting the joint model in JAGS or BUGS using MCMC methods.  This means our focus can move from accounting for the complex sampling process to focusing on specifying biologically meaningful models for the processes of interest, including hierarchical models for the parameters that describe demographic changes in the population.

The house finch example shows how using the CDL facilitates modeling of missing data, including covariate uncertainty.  The model is identical to the one where all data were actually observed, except that we must now consider, and potentially include, the process by which the data go missing.  
This is in contrast to the ODL where any missing data, including covariate uncertainty, usually require specification of a new likelihood.  Examples of this can be seen in \cite{Nichols2004} and \cite{Conn2009} where uncertainty in the covariate requires a new likelihood to be specified in order to include the missing data.

The advantages in using the CDL for modeling comes at a computational cost.  In a Bayesian setting, all missing values are treated as `unknowns', which means each missing value needs to be updated in every MCMC iteration.  Thus the computational burden increases with the amount of missing data.  With current processor power, we are limited in the size of the data sets that we are able to fit.  Using JAGS/BUGS, it can soon become difficult to fit datasets with either a large number of individuals or many sampling periods.  To a large extent, these deficiencies can be overcome with user-written, application specific code, making use of specialized algorithms and other computational advantages  (such as \cite{Dupuis2007} for the multi-state model).  However, this limits generalizability, with new algorithms needed for each application.  These computational limitations, while serious, should not be an impediment to use.  Advances in algorithms for MCMC and continuing increases in computational power mean that we will continue to be able to fit bigger and more complex models into the future.

An issue we have not mentioned is model checking. We suggest that model 
checking be done using posterior predictive checking \cite[][Pg. 159]{Gelman2004}. One approach is to focus on features of the data that are biologically important. For example, we may want to ensure that our 
model explains the difference between the number of individuals caught 
in successive sample occasions. We then generate replicate datasets from 
the fitted model and either (i) visually check, or (ii) specify an 
appropriate test-statistic, to see if the replicates are consistent with 
the observed data. An example of a visual check is shown in \cite[Pg. 164-165]{Gelman2004} with the speed of light data.

An alternative approach is to make use of an omnibus test statistic to 
ensure that our fitted model does not generate data that are 
inconsistent with the data we have observed. Approaches that have been 
adopted in the Bayesian mark-recapture literature include use of the 
likelihood function \cite[]{King2002} or Freeman-Tukey statistic \cite[]{Brooks2000}. 

Goodness-of-fit for mark-recapture models has received considerable 
attention in a frequentist setting 
\cite[]{Pollock1985,Burnham1987,Barker1999, Pradel2003}.  A constructive 
approach to goodness-of-fit testing can be taken when a multinomial 
model is used that is a member of the full exponential family based on 
the factorization
\[
  [\text{Data }|\text{ MSS}] \times [\text{MSS }| \bm \theta].
\]
For a large class of mark-recapture models the term $[\text{Data 
}|\text{ MSS}]$ has a hypergeometric distribution and provides a natural 
partitioning of the data into test components.  As far as we are aware 
an approach based on the predictive distribution $[\text{Data }|\text{ 
MSS}]$ has been little used in a Bayesian setting \cite[see][for an 
exception]{Wright2009} but we believe that this should be a productive 
approach to goodness-of-fit assessment.

A related issue is model selection.  We recommend different approaches to model selection depending on the objective of the study.  If the objective is to learn about the system and to generate hypotheses about relationships then we advocate exploring the data and finding models that best fit the data.  A number of models can be explored and compared using cross-validation \cite[]{Hastie2009} as well as other criteria.  
If the objective is in validating previously generated hypotheses and making inference in the presence of model uncertainty, then we advocate exploring posterior model probabilities, or equivalently Bayes Factors, for a small set of scientifically driven models \cite[]{Link2006}.  \cite{Barker2010} outline an approach to calculating posterior model probabilities that uses the output obtained from running the individual models using MCMC.  Other techniques for calculating posterior model probabilities are described in \cite{King2010}.  Another possible approach to model selection involves specifying a hierarchical model, that has as special cases, all models considered \cite[][Pg. 405 -- 406]{Gelman2004}.  Then, instead of the usual approach of either including the effect or not including the effect before potentially combining the results using model-averaging, we can specify an informative prior distribution centered on zero, that can be viewed as a compromise between inclusion of the effect (with an approximately flat prior) and exclusion of the effect (with a prior with all mass at zero).  An example of this approach is when we have a potential time effect.  Instead of choosing between a model with no time effect, and one with a separate and unrelated parameter for each time point, we can, as we do in the two examples here, include time as a random effect with the variance estimated from the data.

The CDL that we used to fit all models here can be extended in a number of ways.  We expect it to extend naturally to continuous data models.  Depending on the observed data, we would expect the CDL to remain the same, or at least similar, with the conditional likelihood components relating to birth, death, capture and any covariates changing to account for continuous time processes.  

Another extension is when we have uncertainty in the tags themselves.  An example of this is when our ``tag'' is a DNA profile of the individual.  The problem here is that uncertainty in the DNA profile is due to various genotyping errors.  The only change we require in our CDL is to include the true tag as missing data and include a component that describes the corruption of the true tags to the observed tags.  \cite{Wright2009} used this approach to estimate population size in a closed population study.

Using the CDL, we are able to model complex dependencies between individuals in the population.  For example, \cite{Schofield2008} included density dependence on both birth rates and survival probabilities.  Another potential example is one where DNA information for one individual is able to provide information about another individual, such as parents and offspring.  Information about the death of parent gives information about the time of birth of the offspring and vice versa.  The CDL approach is, at least in principle, able to include these relationships as well as including potential uncertainty in the offspring/parent relationship.

\bibliographystyle{asa} 

%

\end{spacing}
\end{document}